\documentclass[useAMS,usenatbib]{mn2e}
\usepackage{psfig, epsf, epsfig}

\title[Optical counterpart of an HLX]{Discovery of an optical counterpart to the hyperluminous X-ray source in ESO\,243-49}
\author[R. Soria et al.]{Roberto Soria$^{1}$\thanks{E-mail:
roberto.soria@mssl.ucl.ac.uk},
George K. T. Hau$^{2}$, Alister W. Graham$^{2}$, Albert K. H. Kong$^{3}$, 
\newauthor
N. Paul M. Kuin$^{1}$, I-Hui Li$^{2}$, Ji-Feng Liu$^{4}$ and Kinwah Wu$^{1}$\\
$^{1}$Mullard Space Science Laboratory, University College London, Holmbury St Mary, Surrey RH5 6NT, UK\\
$^{2}$Centre for Astrophysics and Supercomputing, Swinburne University of Technology, Hawthorn VIC 3122, Australia\\
$^{3}$Institute of Astronomy and Department of Physics, National Tsing Hua University, Hsinchu 30013, Taiwan\\
$^{4}$Harvard-Smithsonian Center for Astrophysics, 60 Garden st, Cambridge MA 02138, USA}

\begin{document}

\date{Accepted 2010 February 09}

\pagerange{\pageref{firstpage}--\pageref{lastpage}} \pubyear{2010}

\maketitle

\label{firstpage}

\begin{abstract}
The existence of black holes of masses $\sim 10^2$--$10^5 {\rm M_{\odot}}$ has important
implications for the formation and evolution of star clusters and supermassive black holes. 
One of the strongest candidates to date is the hyperluminous X-ray source HLX1, possibly located 
in the S0-a galaxy ESO\,243-49, but the lack of an identifiable
optical counterpart had hampered its interpretation. 
Using the {\it Magellan} telescope, we have discovered an unresolved 
optical source with $R = 23.80\pm0.25$ mag and $V = 24.5\pm0.3$ mag within HLX1's 
positional error circle. This implies an average X-ray/optical flux ratio $\sim 500$.  
Taking the same distance as ESO\,243-49, 
we obtain an intrinsic brightness $M_R = -11.0 \pm 0.3$ mag, 
comparable to that of a massive globular cluster.
Alternatively, the optical source is consistent with a main-sequence M star 
in the Galactic halo (for example an M4.4 star at $\approx 2.5$ kpc).
We also examined the properties of ESO\,243-49 by combining {\it Swift}/UVOT
observations with stellar population modelling. We found that 
the overall emission is dominated by a $\sim 5$ Gyr old stellar population, but the
UV emission at $\approx 2000$ \AA\ is mostly due to ongoing star-formation 
at a rate of $\sim 0.03 {\rm M_{\odot}}$ yr$^{-1}$. 
The UV emission is more intense (at least a $9\sigma$ enhancement above the mean) 
North East of the nucleus, in the same quadrant as HLX1.
With the combined optical and X-ray measurements, we put constraints 
on the nature of HLX1. We rule out a foreground star and a background AGN.
Two alternative scenarios are still viable. HLX1 could be an accreting intermediate 
mass black hole in a star cluster, which may itself be the stripped nucleus 
of a dwarf galaxy that passed through ESO\,243-49, an event which might have caused  
the current episode of star formation. Or, it could be a neutron star 
in the Galactic halo, accreting from an M4--M5 donor star.
\end{abstract}


\section{Introduction: ultraluminous and hyperluminous X-ray sources}

{\it XMM-Newton} and {\it Chandra} have discovered several non-nuclear 
X-ray sources in nearby galaxies, with luminosities up to two orders of magnitude higher 
than those observed from Galactic X-ray binaries. These are referred to as  
ultraluminous X-ray sources \citep[ULXs; e.g.,][]{ggs03,swa04,rob07}. 
Those findings have challenged our current models of black hole (BH) formation and accretion.
Isotropic, Eddington-limited luminosities $\ga 10^{40}$erg s$^{-1}$ 
would require BH masses $\ga 100 {\rm M_{\odot}}$, beyond the upper limit for 
individual stellar collapses \citep{yun08}. Mildly super-Eddington luminosity 
(possibly due to large super-Eddington mass accretion) from particularly heavy stellar BHs 
($M \sim 50 {\rm M_{\odot}}$), associated with mildly anisotropic emission  
may explain X-ray luminosities up to $\sim$ few $\times 10^{40}$erg s$^{-1}$ 
without the need for more exotic astrophysical processes \citep{pou07,rob07,kin09}.

Only a few non-nuclear sources have been observed at X-ray luminosities 
$\approx 0.7$--$1 \times 10^{41}$erg s$^{-1}$. For example, in the Cartwheel \citep{wol07}, 
in M\,82 \citep{fen09}, in NGC\,2276 \citep{dav04}, and in NGC\,5775 \citep{li08}. 
It is possible that such rare, extreme ULXs (sometimes known as 
hyperluminous X-ray sources, HLXs) may be powered by heavier BHs, 
formed through different channels: for example, in the collapsed core of 
a super star cluster, or within the nuclear star cluster of an accreted 
(and now disrupted) dwarf galaxy \citep{kin05,bek03}.
Thus, HLXs may represent evidence of intermediate-mass BHs. However, 
the debate is far from settled, given the small number of HLXs known, 
and the possibility of confusion with background AGN.
The strongest claim for an X-ray luminous intermediate-mass BH so far 
has been made for 
a recently discovered X-ray source \citep[2XMM J011028.1$-$460421, hereafter HLX1:][]{far09,god09} 
apparently located in the galaxy ESO\,243-49, or, at least, 
projected inside the $\mu_{B} = 25.0$ mag arcsec$^{-2}$ surface brightness 
isophote of that galaxy.
Here, we report our discovery of 
the likely optical counterpart to this source, and our analysis of the UV emission 
in ESO\,243-49. By determining the optical flux, and the X-ray/optical flux ratio, 
we test alternative models for the nature of this object. Our results strengthen 
the interpretation that the X-ray source belongs to ESO\,243-49. We suggest 
that it is located inside a massive star cluster.

ESO\,243-49 is an edge-on S0-a galaxy at a luminosity distance of $91 \pm 6$ Mpc 
\citep[$z=0.0224$, distance modulus $34.80 \pm 0.15$:][]{cal97}. The foreground 
extinction is very low, $A_V = 0.043$ mag \citep{sch98}.
HLX1 appears projected  
$\approx 7\arcsec$ ($\approx 3.1$ kpc) to the North-East of ESO\,243-49's nucleus, 
and $\approx 1\farcs8$ ($\approx 800$ pc) above the galactic plane.
HLX1 has been detected several times with {\it XMM-Newton}, {\it Chandra} and {\it Swift} 
between 2004 and 2009 \citep{far09,god09}, with an unabsorbed luminosity 
in the $0.3$--$10$ keV band varying between $\la 5 \times 10^{40}$ erg s$^{-1}$ 
and $\approx 1 \times 10^{42}$ erg s$^{-1}$. We also examined a {\it ROSAT}/HRI 
observation of the field from 1996, when HLX1 was not detected to an upper limit 
of $\approx 5 \times 10^{40}$ erg s$^{-1}$. Its combination of extreme luminosity 
(if it really belongs to ESO\,243-49), soft spectrum, 
and spectral changes  
on short timescales \citep{god09}
 makes it a unique object among 
the ULX/HLX class. Its apparent location in an S0 galaxy is also puzzling, because 
such galaxies are usually dominated by an old stellar population. 
For example, the integrated brightnesses of ESO\,243-49 
are (Cousins) $B=14.92 \pm 0.09$ mag, (Cousins) $R=13.48 \pm0.09$ mag and (2MASS) 
$K = 10.70 \pm 0.05$ mag 
(from NED\footnote{http://nedwww.ipac.caltech.edu}), 
which are indicative of a characteristic stellar age $\sim$ a few Gyr (Section 4).
Such moderately old populations were not previously known to host luminous ULXs or HLXs.
For these reasons, it was speculated that 
the source might be a background AGN or a foreground neutron star, even 
though its X-ray properties are also very unusual for both classes of objects (Section 5).

\begin{figure}
\psfig{figure=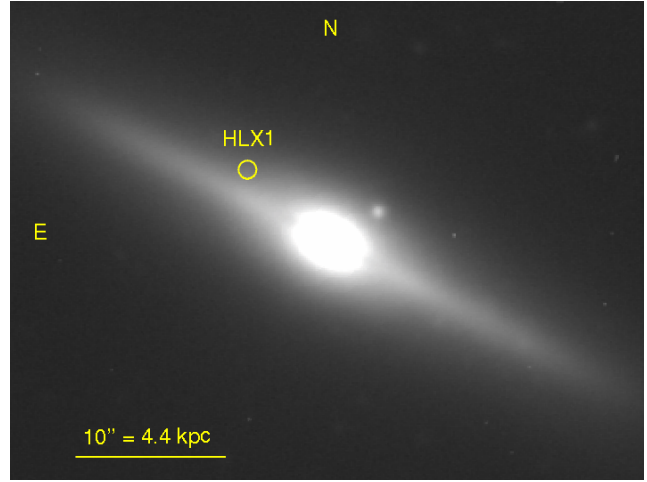,width=84mm}
\caption{{\it Magellan}/IMACS $R$-band image of ESO\,243-49, with the X-ray position of
HLX1 marked by a circle of $0\farcs5$ radius (combined astrometric uncertainty 
of the X-ray and optical images).}
\label{f1}
\end{figure}

\begin{table}
\begin{center}
\begin{tabular}{llrr}
\hline
Telescope & Date  & Band & Exposure \\
\hline
{\it Magellan} Baade & 2009 Aug 26 & R & 540 s  \\
                  &                         & V & 540 s \\
\hline
{\it Swift}/UVOT & 2008 Oct 24 & $u$ &  380 s\\
                  &                         & $uvw1$ & 760 s \\
                  &                         & $uvw2$ & 196 s \\
 & 2008 Oct 25 & $uvw2$ &   1264 s\\
 & 2008 Nov 01 & $u$ &   730 s\\
                  &                         & $uvw1$ & 1690 s \\
                  &                         & $uvw2$ & 2639 s \\
 & 2008 Nov 07  & $u$ &   1210 s\\
                  &                         & $uvw1$ & 2410 s \\
                  &                         & $uvw2$ & 3278 s \\
 & 2008 Nov 08 & $uvw2$ &   582 s\\
 & 2008 Nov 14 & $u$ &   980 s\\
                  &                         & $uvw1$ & 1960 s \\
                  &                         & $uvw2$ & 3814 s \\
 & 2009 Aug 05 & $uvw2$ &  9753 s\\
 & 2009 Aug 06 & $uvw2$ &   9132 s\\
 & 2009 Aug 16 & $uvw2$ &  5664 s\\
 & 2009 Aug 17 & $uvw2$ &   681 s\\
 & 2009 Aug 18 & $uvw2$ &   6032 s\\
 & 2009  Aug 19 & $uvw2$ &   4257 s\\
 & 2009 Aug 20 & $uvw2$ &   2199 s\\
 & 2009  Nov 02 & $uvw2$ & 8956 s\\
 & 2009  Nov 14 & $uvw2$ & 3903 s\\
 & 2009  Nov 20 & $uvw2$ & 517 s\\
 & 2009  Nov 21 & $uvw2$ & 1453 s\\
 & 2009  Nov 28 & $uvw2$ & 2981 s\\
 & 2009  Nov 29 & $uvw2$ & 2028 s\\
 & 2009  Dec 05 & $uvw2$ & 2993   s\\
 & 2009  Dec 19 & $uvw2$ & 2518   s\\
 & 2009  Dec 26 & $uvw2$ & 2774    s\\
 & 2010  Jan 02 & $uvw2$ & 2590   s\\
 & 2010  Jan 08 & $uvw2$ & 4348   s\\
 & 2010  Jan 13 & $uvw2$ & 3577    s\\
 & 2010  Jan 15 & $uvw2$ & 3068    s\\
 & 2010  Jan 22 & $uvw2$ & 2945   s\\
\hline
\end{tabular}
\end{center}
\caption{Optical/UV observation log. }
\end{table}

\begin{figure*}
\psfig{figure=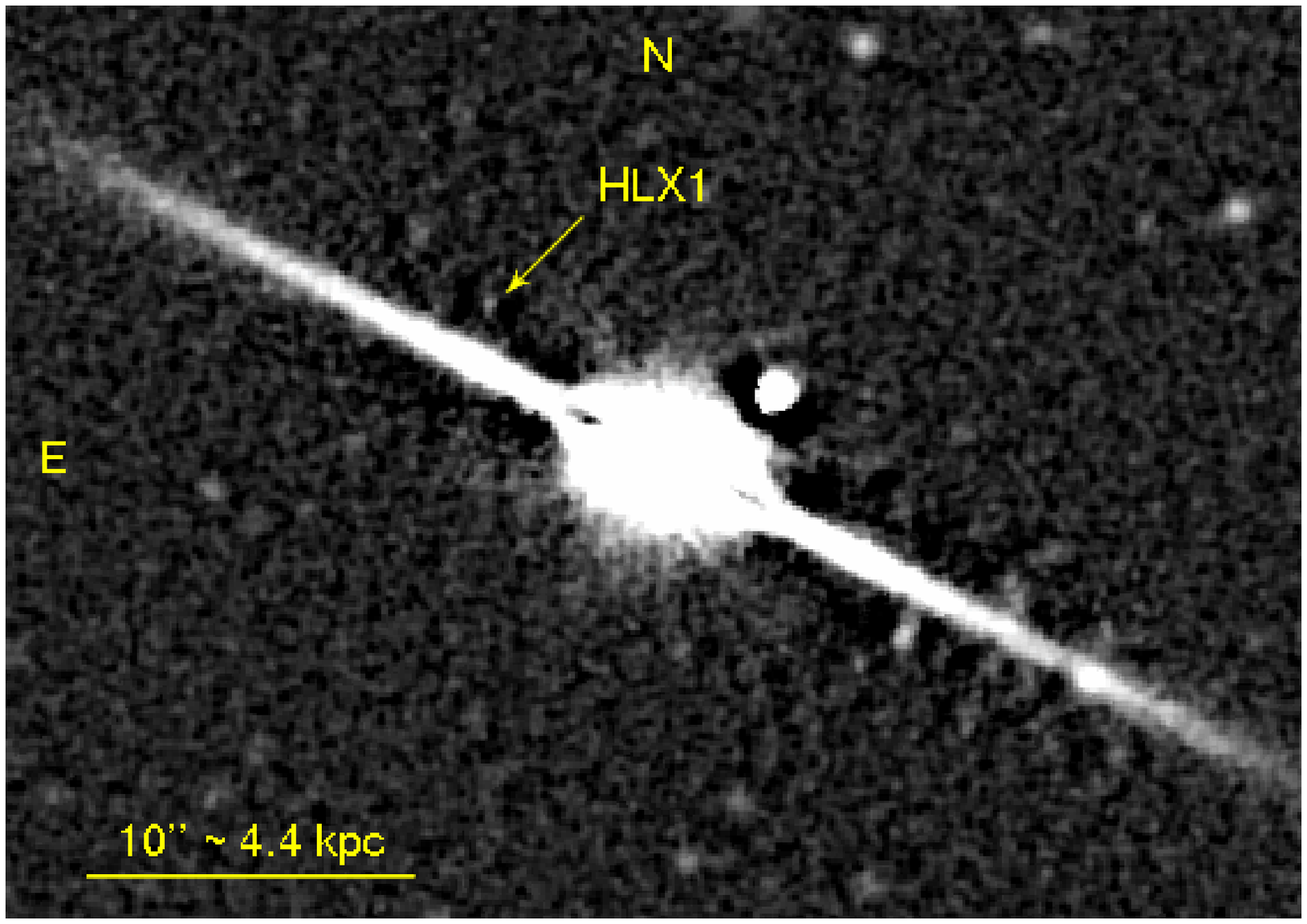,width=87mm}
\psfig{figure=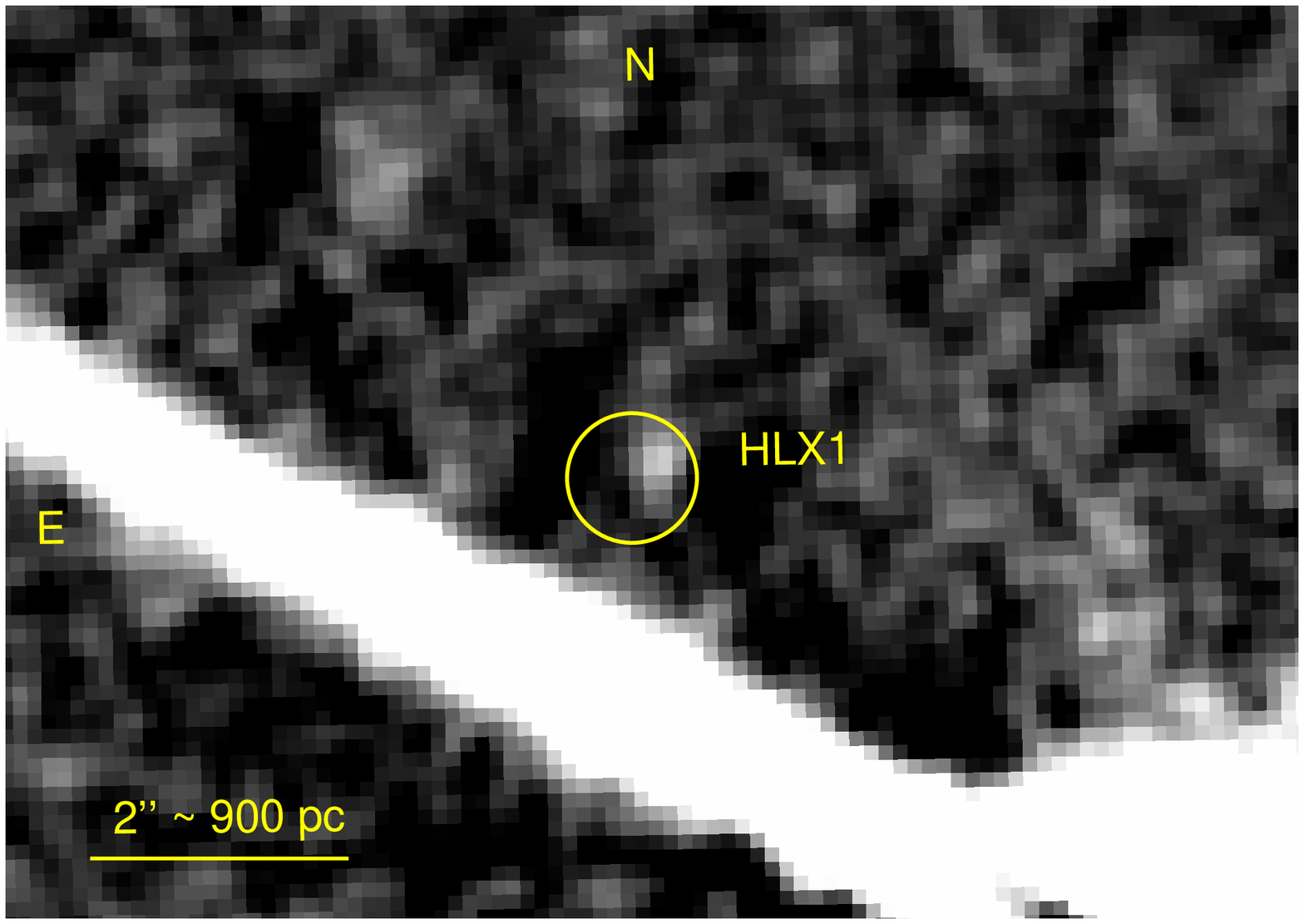,width=87mm}\\
\psfig{figure=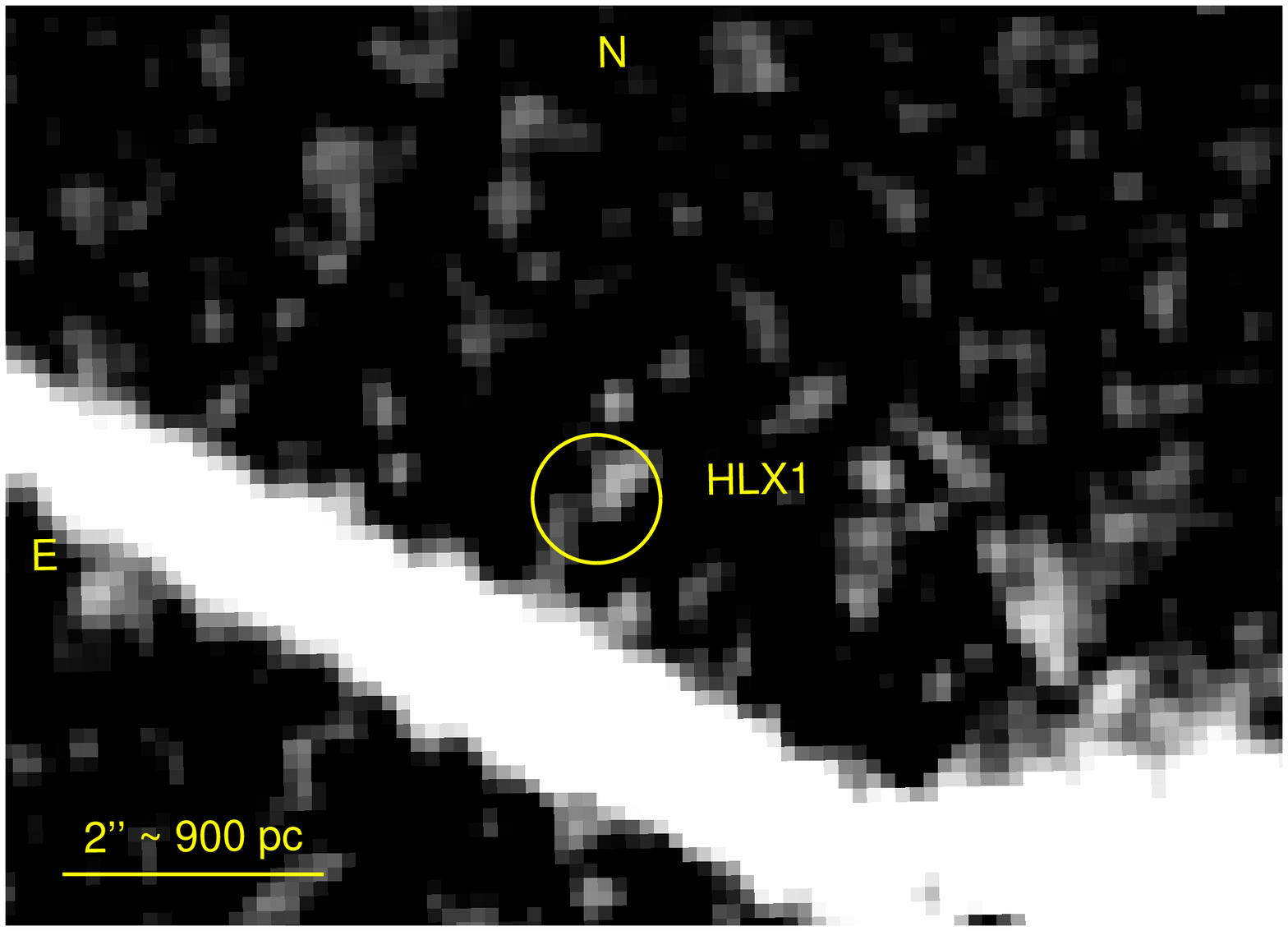,width=87mm}
\psfig{figure=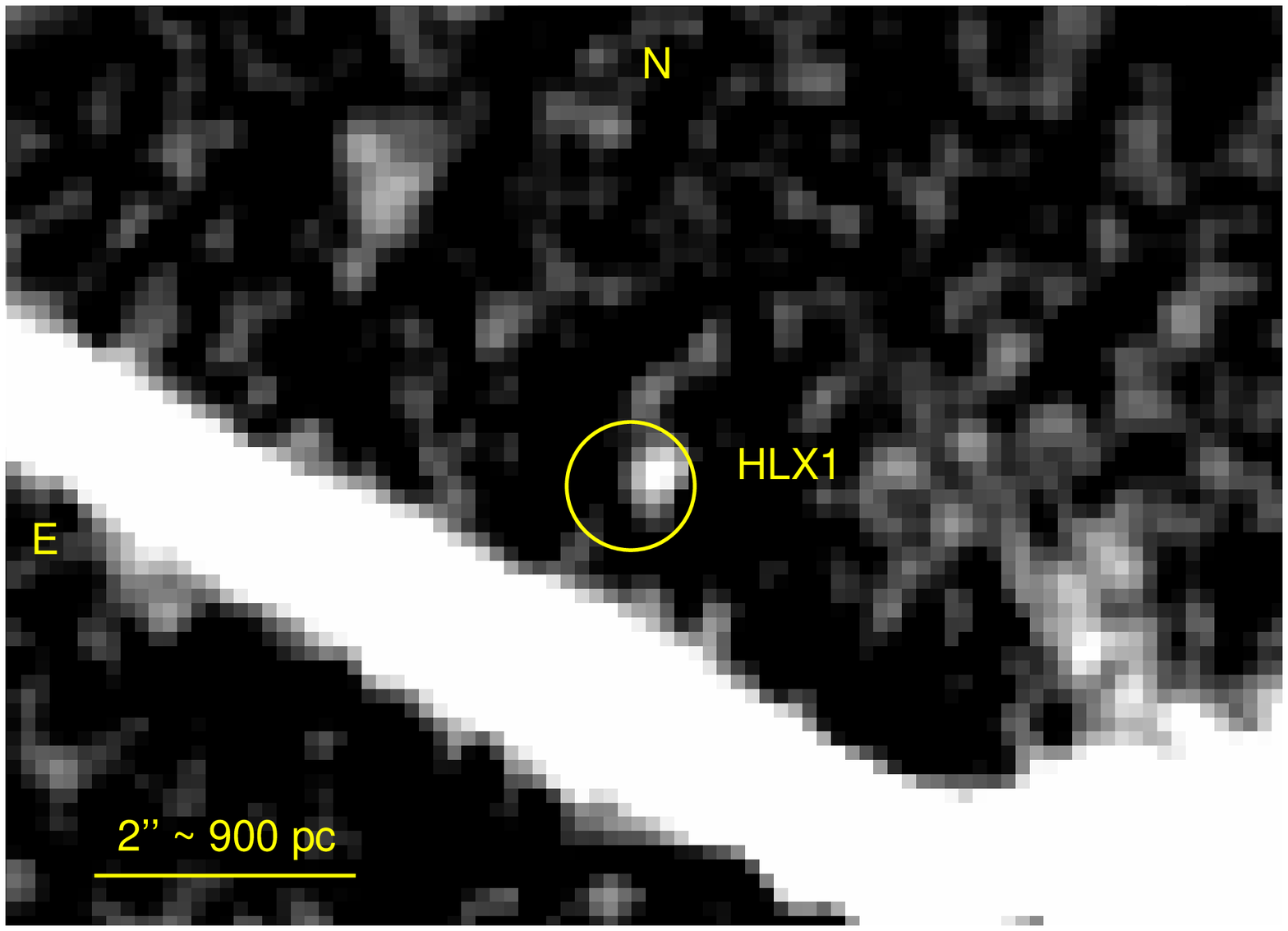,width=87mm}
\caption{{\it Top left:} differential $R$-band image of ESO\,243-49, with a median-filter 
smoothed image subtracted from the original image (logarithmic greyscale). 
The source marked with an arrow has a $4\sigma$ significance and is located 
near the X-ray position of HLX1. 
{\it Top right:} zoomed-in view of the field around HLX1, in the $R$-band residual image 
(square-root greyscale). The {\it Chandra}/HRC-I position of HLX1 
is marked by a circle of $0\farcs5$ radius (combined astrometric uncertainty 
of the X-ray and optical images). Both the top and bottom panel images have been 
Gaussian-smoothed with a kernel radius of 2 pixels ($0\farcs22$), for display purposes only.
{\it Bottom left:} Gaussian-smoothed field around HLX1, in the $V$-band residual image 
(logarithmic greyscale); an optical counterpart is located at the same position 
with a $3\sigma$ significance. {\it Bottom right:} Gaussian-smoothed field around HLX1, 
from the combined $R$-band plus $V$-band residual images.}
\label{f2}
\end{figure*}

\section{X-ray position of HLX1}

We used the {\it Chandra} High Resolution Camera (HRC-I) dataset from 2009 August 17 
(available in the public archives; processed with ASCDVER=8.0) to determine 
the position of the X-ray source HLX1. We checked that there are no processing 
offsets associated with that dataset. Applying any standard source-finding routines 
(e.g., {\it celldetect} or {\it wavdetect} in {\footnotesize{CIAO 4.0}}, 
or {\it imexamine} in {\footnotesize{IRAF}}) we find that HLX1 is a well-isolated, 
on-axis, point-like X-ray source with $\approx 900$ net counts; it is located 
at RA $= 01^h10^m28^s.27$, Dec $= -46^{\circ}04'22\farcs3$, 
with an error radius for the centroid position $\approx 0\farcs03$. We obtain 
excatly the same centroid position when we examine the unbinned HRC-I image 
($0\farcs125$/pixel), or when we bin by 2, or by 4. The 90\% uncertainty circle 
of the HRC-I absolute position 
has a radius of $0\farcs4$\footnote{http://cxc.harvard.edu/cal/ASPECT/celmon/}.
We assessed whether the absolute astrometry of the {\it Chandra}/HRC-I 
image could be improved by using optical/IR/radio counterparts 
with well-known positions. There are only two other (fainter) X-ray sources 
with $> 25$ counts located within $\approx 5\arcmin$ of the aimpoint; 
neither has a known counterpart. A few other X-ray sources are located farther 
from the aim point, and therefore have a very elongated PSF and are not good choices 
for astrometry calibration. 
We also tried registering the three {\it Chandra} sources nearest to the aimpoint 
onto the corresponding {\it XMM-Newton} sources, and then some of the other 
{\it XMM-Newton} sources onto their optical/IR counterparts, but we concluded that 
this does not improve either the precision or the accuracy of the 
original {\it Chandra} astrometry.


\begin{table*}
\begin{center}
\begin{tabular}{lrrrrr}
\hline
UVOT Filter  & Brightness  & $\lambda_{\rm eff}$ & $\nu_{\rm eff}$ & Flux Density  & Flux Density\\
 & (mag) & (\AA) & ($10^{15}$ Hz) & ($10^{-16}$ erg s$^{-1}$ cm$^{-2}$ \AA$^{-1}$) 
& ($10^{-26}$ erg s$^{-1}$ cm$^{-2}$ Hz$^{-1}$)\\
\hline
$uvw2$ & $18.06 \pm 0.12$ &  $2030$ &  $1.477$ & $3.1 \pm 0.3$ & $0.045 \pm 0.005$\\
$uvw1$ & $17.07 \pm 0.12$ &  $2634$ &  $1.138$ & $6.3 \pm 0.6$ & $0.154 \pm 0.015$\\
$u$ & $15.7 \pm 0.2$ &  $3501$ &  $0.856$ & $16.6 \pm 2.4$ & $0.85 \pm 0.12$\\
\hline
\end{tabular}
\end{center}
\caption{Integrated brightness of ESO\,243-49 in the UVOT bands ($3\sigma$ confidence level). }
\end{table*}

\section{Optical counterpart of HLX1}

We observed the source on 2009 August 26 with the IMACS Long Camera (SITe CCD) 
on the 6.5-m Baade {\it Magellan} telescope. We took a series of $3 \times 180$ s exposures 
in each of the Bessell {\it B,V,R} filters \citep{bes79}. However, 
the {\it B} images are badly affected by a non-optimal focus 
and cannot be used for this work. The seeing was $\approx 0\farcs7$, 
and the airmass was $\approx 1.2$.
We used the {\footnotesize {STARLINK}} program {\tt gaia}, as well as 
standard {\footnotesize {IRAF}} packages, to analyze the optical images.
We calibrated the astrometry of the {\it Magellan} images by using 
stellar positions from the Guide Star Catalog Version 2.3 \citep{las08}.
Our optical astrometry is accurate to $\approx 0\farcs3$.

At first inspection, any faint optical counterpart to the X-ray source 
would be swamped by the strong stellar light from the main galaxy (Figure 1).
However, at closer inspection we noted that there is an excess 
of counts, consistent with a point source, in the $R$-band image 
at the position of HLX1. To visualize this source, we generated
a smoothed image with a $17 \times 17$ pixel median filtering. 
We then subtracted the smoothed image from the original image.
This brings up residuals consistent with point-like sources, 
including a source in the X-ray error circle of HLX1 (Figure 2, 
top panels). 
The optical source is located at RA $= 01^h10^m28^s.25$, Dec $= -46^{\circ}04'22\farcs2$, 
which is $\la 0\farcs3$ from the central X-ray position, 
and within the combined optical and X-ray positional uncertainties ($\approx 0\farcs5$).
There are no other optical sources of comparable or higher brightness within 
a few arcsec, outside the galactic plane.
As a further check that the optical source is not for example a cosmic ray, 
we analysed each of the three 180-s sub-exposures separately, 
and found it in each of them.
To quantify the significance of this detection, we measured the counts 
in a $5 \times 5$ pixel box centred on the source in the residual image 
(multiplied by an aperture correction factor of 1.1, to account 
for the small fraction of source counts falling outside the box).
We compared those counts with the mean and the distribution of counts 
in 20 $5 \times 5$ pixel boxes covering the background sky around HLX1 
(above and below the disk plane of the galaxy). 
We obtain that the source detection is significant to $4\sigma$.
Moreover, here we are not looking for any $4\sigma$ source randomly 
positioned around the galaxy: we are specifically investigating whether 
there is a source within $\approx 0\farcs5$ of the X-ray position. 
Based on the source significance combined with the positional coincidence, 
we conclude that this optical source is real and is the most likely candidate 
for the optical counterpart of HLX1.
We then repeated the same analysis for the $V$-band image, 
and found a source at $3\sigma$ significance in the residual image, 
at the same position as the source in the $R$-band image (Figure 2, 
bottom left panel). The postional coincidence further strengthens 
our identification as the likely optical counterpart of HLX1. 
Finally, we combined the $R$ and $V$ residual images, which 
further improves the signal-to-noise ratio (Figure 2, 
bottom right).

We performed an aperture photometry measurement of this source, and used 
isolated point-like objects in the field to calculate the aperture correction. 
To convert from count rates to fluxes, at first we tried using 
stars with photometric measurements in the Naval Observatory Merged Astrometric 
Dataset (NOMAD) Catalog \citep{mon03,zac05}. However, we did not find 
stars with sufficiently reliable and accurate brightnesses within 
the field of view of the {\it Magellan} image. We then downloaded 
a series of $V$ and $R$ exposures taken with the Wide Field Imager (WFI) 
on the 2.2-m MPG/ESO telescope at La Silla\footnote{The WFI images were taken 
as part of a programme by H. Boehringer, to look for high-redshift clusters.}, 
which obviously cover a much larger field of view around ESO\,243-49.
We measured the brightness of a few galaxies with reliable NED 
values to calibrate the photometry of the WFI field, and applied it 
to sources that are also included in the {\it Magellan} images. 
From that, we bootstrapped the conversion between count rate and magnitudes 
in the {\it Magellan} images. For ESO\,243-49, we obtain an integrated 
(Cousin) brightness $R = 13.5 \pm 0.1$ mag, $V = 14.2 \pm 0.1$. This is 
in agreement with the values listed in NED, and with the expected colour 
of a moderately old population; it shows that our method 
is reliable\footnote{As a further check, we also examined the count rate to magnitude 
conversion factors from the {\it Magellan} exposure time calculator, 
and found that they agree with our conversion factors, within $\approx 0.3$ mag.}.
Using the same photometric calibration and accounting 
for the aperture correction of a point-like source, we determined 
the brightness of the optical counterpart to HLX1.
We obtain a brightness $R = 23.80 \pm 0.25$ mag, $V = 24.5 \pm 0.3$ mag.
If the object is located in ESO\,243-49, its intrinsic brightness (corrected 
for Galactic extinction and taking into account the distance uncertainty) 
is $M_R = -11.0 \pm 0.3$ mag, $M_V = -10.4 \pm 0.3$ mag.

\begin{figure}
\psfig{figure=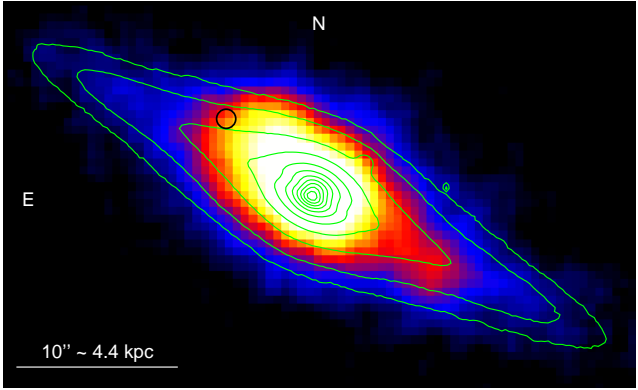,width=84mm}
\caption{{\it Swift}/UVOT false-colour image of ESO\,243-49 in the $uvw2$ band, 
with $R$-band surface brightness contours overplotted (square root scale).
The X-ray position of HLX1 is marked with a black circle (error radius $0\farcs6$: 
combined astrometric uncertainty of the X-ray and UV images).
While the $R$-band contours are highly symmetric, the $uvw2$ emission extends 
more strongly to the North East of the nucleus, with a $>9\sigma$ significance 
of the excess. The flux density scale corresponding to the $uvw2$ false-colours 
is as such: 
blue between $\approx 0.2$--$0.7 \times 10^{-18}$ erg s$^{-1}$ cm$^{-2}$ \AA$^{-1}$ arcsec$^{-2}$ 
($uvw2 \approx 26$--$24.7$ mag arcsec$^{-2}$); 
red between $\approx 0.7$--$1.4 \times 10^{-18}$ erg s$^{-1}$ cm$^{-2}$ \AA$^{-1}$ arcsec$^{-2}$ 
($uvw2 \approx 24.7$--$23.9$ mag arcsec$^{-2}$); 
yellow between $\approx 1.4$--$2.1 \times 10^{-18}$ erg s$^{-1}$ cm$^{-2}$ \AA$^{-1}$ arcsec$^{-2}$ 
($uvw2 \approx 23.9$--$23.5$ mag arcsec$^{-2}$); 
white $\ga 2.1 \times 10^{-18}$ erg s$^{-1}$ cm$^{-2}$ \AA$^{-1}$ arcsec$^{-2}$ 
($uvw2 \la 23.5$ mag arcsec$^{-2}$).
}
\label{f3}
\end{figure}

\section{UV emission from ESO\,243-49}

In the UV bands, ESO\,243-49 was observed several 
times between 2008 October 24 and 2010 January 22, with the 30-cm UV/Optical 
Telescope (UVOT) on board {\it Swift}. The total exposure times were 3.3 ks 
in the {\it u} band, 6.7 ks in {\it uvw1}, 4.9 ks in {\it uvm2}, and 94.1 ks in {\it uvw2}. 
In fact, the {\it uvm2} exposure proved to be too short for meaningful analysis, 
and we do not use it further. See Table 1 for a log of our observations.
We retrieved the {\it Swift}/UVOT datasets (all observations  
through the end of 2009 November) from the HEASARC archive, and  
processed them using standard {\footnotesize {FTOOLS}} tasks.
We applied the mod8 and aspect corrections to each of the individual images, 
which were taken at various space craft roll angles, and with slightly 
different offsets. Most images were provided with a $1\arcsec$ pixel size. 
We summed them using the {\tt uvotimsum} task, and resampled them 
to a $0\farcs5$ pixel size.
We checked the astrometry by correlating UVOT sources with those detected 
in the Magellan images, as well as source positions from the USNO-B1.0 Catalog \citep{mon03}.
In particular, we determined the position error between 23 sources common to 
the $R$-band and the combined {\it uvw2} image, and found that the positions 
are consistent with a root-mean-square spread of $0\farcs18$, with 
a 95\% confidence level of $\approx 0\farcs4$.
We conservatively took a combined error circle of radius $0\farcs6$ 
when comparing UVOT and {\it Chandra} positions.
The observed full-width half-maximum of point-like sources 
in the combined images goes from $\approx 3\farcs0$ for the {\it u} filter image to 
$\approx 3\farcs6$ for the {\it uvw2} filter image, which is a stack 
of 90 sub-exposures; the observed full-width half-maximum in each 
{\it uvw2} subexposure is $\approx 3\farcs4$. 
We performed aperture photometry with {\tt uvotsource} to derive 
initial count rates. Filter transmission curves and conversion factors between 
UVOT count rates and fluxes or magnitudes are detailed in \citet{poole}. 
However, those factors were derived for an aperture of $5\arcsec$ on point sources.  
In our case, we are also interested in the total emission from 
extended regions. Therefore, we calculated appropriate 
aperture corrections for extended emission, 
using the extended UVOT point-spread-function 
\citep{breeveld}, and the appropriate HEASARC {\it Swift}/UVOT CALDB files.

We do not detect a point-like UV counterpart at the X-ray position, 
to an upper limit $uvw2 \approx 22.7$ mag; this corresponds to a flux 
$f_{\nu} \approx 5.9 \times 10^{-7}$ Jy $= 5.9 \times 10^{-30}$erg s$^{-1}$ cm$^{-2}$ Hz$^{-1}$ 
at an effective frequency of 
$1.477 \times 10^{15}$ Hz, or $\approx 2030$ \AA\ \citep{poole}.
The non-detection of a point-like counterpart is not surprising, 
given the full-width half-maximum of the point spread function. 
However, the UVOT data provide interesting information on 
the host galaxy environment. 
For the whole galaxy, we measure a total brightness (uncorrected for extinction) 
$u = 15.7 \pm 0.2$ mag\footnote{The difference between the UVOT $u$ brightness 
and the standard Johnson U is $\la 0.05$ mag \citep{poole}.}, 
$uvw1 = 17.07 \pm 0.12$ mag, $uvw2 = 18.06 \pm 0.12$ mag 
($3\sigma$ uncertainties). The $uvw2$ brightness corresponds to 
a flux $f_{\nu} \approx 4.5 \times 10^{-5}$ Jy $= 4.5 \times 10^{-28}$erg s$^{-1}$ cm$^{-2}$ Hz$^{-1}$ 
at $\nu = 1.477 \times 10^{15}$ Hz (Table 2).

First, we tested whether these UV brightnesses and colours are consistent 
with the moderately old population suggested by the optical colours and morphology 
of ESO\,243-49.
We downloaded\footnote{From http://www-wfau.roe.ac.uk/6dFGS} 
an optical spectrum of the galaxy from the 6dF Galaxy 
Survey Database \citep{jon04,jon09}.
From the observed strength of the H$\beta$ 
absorption line, and of the Fe5270 and Fe5335 indices \citep{wor04a,wor04b}, 
we estimate that the dominant population has an age $\approx 4.5^{+4.0}_{-2.5}$ Gyr 
(assuming solar metallicity).
As a further check of this observational estimate, we ran instantaneous 
star-formation simulations with Starburst99 \citep{lei99,vaz05}, to determine 
the expected brightness and colours of stellar populations with ages $\sim 2$ to $8$ Gyr.
In particular, we find that a population with a single age of 4.5 Gyr, 
initial stellar mass $= 6 \times 10^{10} M_{\odot}$ and solar metallicity
is predicted to have 
$R \approx 13.5$ mag, $B \approx 15.0$ mag, $U \approx 15.5$ mag, 
$uvw1 \approx 17.3$ mag, $uvw2 \approx 20$ mag (at the distance 
of ESO\,243-49 and after adding the foreground extinction 
$A_V = 0.043$ mag). Such values are indeed similar to the observed 
colours, except for $uvw2$. More detailed population modelling 
of the galaxy is beyond the scope of this work: here, we simply 
want to stress that the emission in all bands 
up to $uvw1$ (effective wavelength $2634$ \AA) is dominated 
by a moderately old stellar population, but extra UV emission 
from a much younger population dominates the $uvw2$ band. 
Adding a population with ongoing star-formation rate 
$\approx 0.03~{\rm M_{\odot}}$ yr$^{-1}$ 
is sufficient to explain the bright far-UV emission, 
while it contributes little to the other bands, 
compared to the emission from the older population (Table 3).

\begin{table*}
\begin{center}
\begin{tabular}{lrrrrrr}
\hline
Band  & Obs. Brightness  & Obs. Flux Density & Model Flux Density & Model Flux Density  
       & Model Brightness & Model Brightness \\
 &       &                  & Old Population & Young Population & Old Population & Combined \\
 & (mag) & ($10^{-16}$ CGS) & ($10^{-16}$ CGS) 
& ($10^{-16}$ CGS) & (mag) & (mag)\\
\hline
$uvw2$ & $18.06 \pm 0.12$ & $3.1 \pm 0.3$ & $\approx 0.5$ & $\approx 2.5$ & $\approx 20$ & $\approx 18.1$\\
$uvw1$ & $17.07 \pm 0.12$ & $6.3 \pm 0.6$ & $\approx 5$ & $\approx 1.1$ & $\approx 17.3$ & $\approx 17.1$\\
$u$ & $15.7 \pm 0.2$ & $16.6 \pm 2.4$ & $\approx 19$ & $\approx 0.6$ & $\approx 15.5$ & $\approx 15.5$\\
\hline
\end{tabular}
\end{center}
\caption{Observed brightness of ESO\,243-49 in the UVOT bands, compared with 
the predicted brightness of a 4.5-Gyr-old population (initial stellar mass 
$= 6 \times 10^{10} M_{\odot}$, solar metallicity, foreground extinction 
$A_V = 0.043$ mag) with an additional contribution 
from ongoing star formation at a rate $\approx 0.03~{\rm M_{\odot}}$ yr$^{-1}$. 
Flux units are $10^{-16}$ erg s$^{-1}$ cm$^{-2}$ \AA$^{-1}$.
We used Starburst99 \citep{lei99,vaz05} for the model simulations.}
\end{table*}

We then examined the emission from the young stellar population, 
which dominates the {\it uvw2} image (Figure 3).
The emission appears asymmetric, and does not match 
the $R$-band surface brightness contours well. In particular, it is clear 
already from an eye-ball inspection that the
{\it uvw2} emission extends more strongly to the North East of the nucleus 
(the same quadrant as HLX1). 
In order to quantify the degree of asymmetry, we used 
the highly symmetric $R$-band isophotes to define four quadrants 
(bounded by the galactic plane and the normal to the plane), 
and extracted the {\it uvw2} flux in each of them. 
Based on our best optical/UV alignment, we find that the flux in the North East 
quadrant is $\approx 40$\% 
higher than the average flux of the other three quadrants; that corresponds 
to a $13\sigma$ enhancement. Excluding a $4\arcsec$-box around the nucleus, 
we measured the following {\it uvw2} net count rates in the four quadrants. 
North-East: $0.1100 \pm 0.0017$ counts s$^{-1}$; 
North-West: $0.0784 \pm 0.0016$ counts s$^{-1}$;
South-West: $0.0792 \pm 0.0016$ counts s$^{-1}$;
South-East: $0.0794 \pm 0.0016$ counts s$^{-1}$.
We tried different definitions for the size of the four quadrants, 
including/excluding the nucleus. We repeated the exercise after shifting 
the placement of the four quadrants by $\pm 0\farcs4$, to account 
for the relative astrometric uncertainty between the UVOT and Magellan images 
and the small uncertainty in the position angle of the major axis. 
We obtain that the emission enhancement in the North East 
quadrant is always significant to $\ga 9\sigma$.
The excess {\it uvw2} emission in that part of the galaxy   
may be interpreted as a more recent or intense phase 
of star formation.
We find no statistically-significant enhancements or asymmetries 
in the {\it uvw1} and {\it u} bands, in agreement with our estimate 
that the emission in those bands is dominated by the old stellar population.
We can plausibly say that the young/starforming 
component is not as symmetrically distributed as the old population.

\section{Discussion}

If the X-ray source HLX1 is proven to be an accreting BH with 
mass $\sim 10^3$--$10^4 {\rm M_{\odot}}$,
there would be important implications on models of galaxy formation 
and evolution. Identifying its optical counterpart gives a crucial constraint 
on its nature. We have found an unresolved optical source within its X-ray error circle, 
and it is likely to be physically associated to HLX1. 
Assuming a direct association, we calculate an X-ray/optical flux ratio, 
using the standard definition \citep{hor01}:
$\log(f_{\rm X}/f_{\rm R}) = \log f_{\rm X} + 5.5 + R/2.5$, 
where $f_{\rm X}$ is the intrinsic flux in the $0.3$--$10$ keV band  
\citep[taken from][and our spectral analysis 
of the {\it Swift}/XRT data]{god09}. 
We obtain $f_{\rm X}/f_{\rm R} \approx 800$--$1000$ for the X-ray high state 
of 2009 August, and $f_{\rm X}/f_{\rm R} \approx 500$ 
for the ``average'' X-ray state where the source 
was more often observed over 2008--2009; such ratio 
is only slightly dependent on the choice of X-ray spectral model.
%
The observed X-ray/optical flux ratio is much higher 
than expected for AGN, which have typical  
$f_{\rm X}/f_{\rm R} \la 10$ \citep{lai09,bau04}. A number of distant, faint AGN are undetected 
in the optical band because of extinction, which is not an issue 
for this object \citep{far09,god09}. 
In particular, AGN with a $0.5$--$2$ keV flux of $\approx 5 \times 10^{-13}$ 
erg s$^{-1}$ cm$^{-2}$ are rare \citep[$N \approx 0.5$ deg$^{-2}$:][]{has98} 
and have always been easily identified in other bands. 
Moreover, the red colour of the optical counterpart ($V-R = 0.7 \pm 0.4$) 
suggests that the optical emission is not dominated by the Rayleigh-Jeans 
tail of an accretion disk spectrum.

Neutron stars are a class of objects that can reach X-ray/optical flux ratios $\ga 1000$, 
with thermal X-ray emission from their surface. 
We note (Soria et al., in prep.) that the X-ray spectra of HLX1 
can also be fitted with a neutron star atmosphere model (e.g., {\tt nsa} 
in {\footnotesize XSPEC}) plus power-law.  
Such models suggest a characteristic distance $\approx 1.5$--$2.5$ kpc, 
placing the source in the Galactic halo. The corresponding 
$0.3$--$10$ keV luminosity would be $\approx$ a few $\times 10^{32}$ erg s$^{-1}$. 
Intriguingly, this is a range of luminosities where quiescent 
low-mass X-ray binaries also show a thermal plus power-law spectrum, 
with the relative contribution of the power-law decreasing 
as the source gets brighter \citep{jon04}.
In this scenario, the apparent optical brightness $R \approx 23.8$ mag implies 
$M_R \approx 11.8$--$12.8$ mag, consistent with a main-sequence M4.4--M5.2 donor star, 
with an initial mass $\approx 0.13$--$0.17 {\rm M_{\odot}}$ \citep{kni06,gir00}. 
A late-type M star is also consistent 
with the observed red colour. Thus, we cannot rule out 
the possibility of an accreting neutron star in the Galactic halo, 
from the optical data. In this case, the excess UV emission North East of the nucleus 
in ESO\,243-49 is purely coincidental.

The optical counterpart of HLX1 does not stick out like a unique object 
in the field. In the $R$-band residual image, we identified a few other sources consistent 
with globular clusters around ESO\,243-49, with comparable brightnesses (Figure 2, top panel).
If the HLX1 counterpart is also a globular cluster in that galaxy, 
its optical luminosity would place it between the Milky Way globular cluster 
$\omega$ Cen \citep[$M_V = -10.3$ mag, $M_R = -10.8$ mag, 
$M_{\rm tot} \approx 2.8 \times 10^6 {\rm M_{\odot}}$:][]{har96,van09}
and the Andromeda cluster G1 
\citep[$M_V = -11.2$ mag, 
$M_R = -11.8$ mag, $M_{\rm tot} \approx 5 \times 10^6 {\rm M_{\odot}}$:][]{gra09}, 
which is a strong candidate for the presence 
of an $\approx 2 \times 10^4 {\rm M_{\odot}}$ BH \citep[][but see \citealt{bau03}]{geb05}.

There are a number of scenarios for the formation 
of an intermediate-mass BH inside a massive star cluster. 
In a young star cluster, runaway core collapse and coalescence of the most 
massive stars can occur over a timescale $\la 3$ Myrs 
and can result in the formation of a supermassive star, 
which can quickly collapse into a BH \citep{por02,fre06}.
In an old globular cluster, an intermediate-mass BH 
can be formed from the merger of stellar-mass BHs and neutron stars  
over a timescale $\sim 10^9$ yr \citep{ole06}.
Subsequent capture and disruption of a cluster star 
may then provide the accretion rate required to explain 
the X-ray luminosity. In this scenario, HLX1 and its host 
star cluster are unrelated to the ongoing star formation 
in the bulge of ESO\,243-49.

The most intriguing scenario is that some massive star clusters may have 
been the nuclear clusters of satellite galaxies accreted 
and tidally disrupted by a more massive galaxy. 
Dwarf galaxies are the most common type of galaxies 
in clusters \citep[e.g.,][]{bin85} and many of them 
are nucleated \citep[e.g.,][]{gra03,cot06}. In many cases, 
a nuclear cluster may coexist with a nuclear BH \citep{gra09,set08}. 
This may end up in the halo of a bigger galaxy after a merger. 
$\omega$ Cen itself may have originated from the nuclear star cluster 
of an accreted dwarf \citep{bek03}. Similar suggestions have been 
made for a group of clusters in NGC\,5128 \citep{pen02,cha09}.
The recent or ongoing star formation in ESO\,243-49 may have been triggered 
by the passage and tidal disruption of the satellite galaxy, 
perhaps along the South-West to North-East direction, since 
{\it uvw2} emission is stronger on that side (Section 4).
During its passage through ESO\,243-49, the compact nucleus of the satellite dwarf 
may also have collected gas from the main galaxy \citep{pfl09}, and this may perhaps 
be fuelling a nuclear BH. In this scenario, HLX1 may be the intermediate-mass BH 
located in the nuclear cluster of that accreted satellite.

In summary, we have identified a point-like optical counterpart 
for HLX1 in ESO\,243-49 in the {\it Magellan} images.
The optical brightness and colour are consistent with a massive star cluster 
in ESO\,243-49, or with a main-sequence M4--M5 star in the Milky Way halo, 
at $\approx 1.5$--$2.5$ kpc. The galaxy is dominated by a $\sim 5$ Gyr old population, but 
shows excess emission in the {\it Swift}/UVOT $uvw2$ band, consistent 
with a recent episode of star formation. The far-UV emission has an asymmetric shape 
and is stronger to the North East of the nucleus, roughly in the direction of HLX1. 

\section*{Acknowledgments}

We thank Mark Cropper, Rosanne di Stefano, Sean Farrell, Jeanette 
Gladstone, Craig Heinke, Erik Hooverstein, Pavel Kroupa, Tom Maccarone, Greg Sivakoff, 
Lee Spitler and Doug Swartz for discussions. We thank the referee for 
a detailed review and comments. RS acknowledges hospitality 
and financial assistance at Tsing Hua University (Taiwan), 
and hospitality at the University of Sydney, during part of this work.

\end{document}